\documentstyle[12pt,psfig,draft]{nature}

\def\simlt{\mathrel{\hbox{\rlap{\hbox{\lower4pt\hbox{$\sim$}}}\hbox{$<$}}}}
\def\simgt{\mathrel{\hbox{\rlap{\hbox{\lower4pt\hbox{$\sim$}}}\hbox{$>$}}}}
    
\begin{document}


\title{\Large \bf BEAMING IN GAMMA-RAY BURSTS: \hfil\eject
EVIDENCE FOR A STANDARD ENERGY RESERVOIR}
 
\author{
D. A. Frail\affiliation[1]{National Radio Astronomy Observatory, P.O. Box
    0, Socorro, NM 87801, USA},
  S. R. Kulkarni\affiliation[2]{California Institute of Technology,
    Palomar Observatory 105-24, Pasadena, CA
    91125, USA}
\affiliation[3]{Massachusetts Institute of Technology,
    Physics Dept., Cambridge, MA 02139, USA},
R. Sari,\affiliation[4]{California Institute of Technology, Theoretical
 Astrophysics 130-33, Pasadena, CA 91125, USA}
S. G. Djorgovski$^\dagger$,
J. S. Bloom$^\dagger$,
T. J. Galama$^\dagger$,
D. E. Reichart$^\dagger$,
E.    Berger$^\dagger$,
F. A. Harrison$^\dagger$,
P. A. Price$^\dagger$\affiliation[5]{School of Astronomy \&
  Astrophysics, Mount Stromlo Observatory, Cotter Road, Weston, ACT,
  2611, Australia},
S. A. Yost$^\dagger$,
A. Diercks$^\dagger$,
R. W. Goodrich\affiliation[6]{W. M. Keck Observatory, Kamuela, HI
  96743, USA},
F. Chaffee$^\parallel$
}

\dates{\today}{} 
\headertitle{}
\mainauthor{}
\smallskip  

\summary{ Gamma-ray bursts (GRBs) are the most brilliant objects in
  the Universe but efforts to estimate the total energy released in
  the explosion -- a crucial physical quantity -- have been stymied by
  their unknown geometry: spheres or cones.  We report on a
  comprehensive analysis of GRB afterglows and derive their conical
  opening angles. We find that the gamma-ray energy release, corrected
  for geometry, is narrowly clustered around $5\times 10^{50}$ erg. We
  draw three conclusions. First, the central engines of GRBs release
  energies that are comparable to ordinary supernovae, suggesting a
  connection.  Second, the wide variation in fluence and luminosity of
  GRBs is due entirely to a distribution of opening angles.  Third,
  only a small fraction of GRBs are visible to a given observer and
  the true GRB rate is at least a factor of 500 times larger than the
  observed rate.  }

\maketitle

\noindent{\it This paper has been submitted to Nature.
  You are free to use the results here for the purpose of your
  research.  In accordance with the editorial policy of Nature, we
  request that you not discuss this result in the popular press. If
  you have any question or need clarifications please contact Dale
  Frail, dfrail@nrao.edu or Shri Kulkarni, srk@astro.caltech.edu.}

\newpage

\noindent Observations of GRBs are well described by the fireball 
model\cite{piran00}, in which an explosive flow of relativistic matter
(ejecta) is released from a central source.  The collision of
fast-moving ejecta with slower moving ejecta result in bursts of gamma
rays.  Shortly thereafter, the ejecta starts shocking and sweeping up
significant amounts of circumburst matter.  The shocked gas, hereafter
called the blast wave, powers long-lived and broad-band (X-ray,
optical and radio) emission -- the so-called afterglow emission.

The afterglow emission appears to be primarily synchrotron radiation.
As with other astrophysical shocks, the shocked electrons are
accelerated to relativistic energies, forming a power-law
distribution, $dN/dE_e \propto E_e^{-p}$; here, $E_e$ is the energy of
the electron and $p$, the index of the power law. In the presence of
magnetic fields, the electrons radiate synchrotron emission with a
flux $f(t,\nu)\propto t^\alpha\nu^\beta$, where the spectral index
($\beta$) and the temporal index ($\alpha$) are related to $p$ and the
dynamics of the blast wave\cite{spn98}. Broad-band observations have
repeatedly confirmed the expectations of this simple
picture\cite{wax97a,gwb+98,bsf+00}.

The two outstanding issues in this field are (a) determining the
progenitors of GRBs and (b) understanding the physics of the central
engine. The focus of this {\it Article} is the latter topic,
specifically the energetics of these mysterious sources.

Observationally it is known\cite{fac+00} that the fluence (defined as
the received energy per unit area) of the broad-band afterglow phase
is always (and usually much) smaller than that of the gamma-ray burst
fluence.  This then motivates the use of the isotropic equivalent
gamma-ray energy, $E_{\rm iso}(\gamma)= 4\pi F_\gamma
d_L^2(1+z)^{-1}$ as a surrogate for the energy released by the central
engine. Here, $F_\gamma$ is the fluence of the burst; $z$ is the
redshift; and $d_L$ is the luminosity distance.

However, $E_{\rm iso}(\gamma)$ could grossly overstate the true
gamma-ray energy release ($E_\gamma$) if the explosion is not
spherical. Indeed, jets are present in almost all accretion-driven
phenomena, e.g., young stellar objects, neutron star binaries, and
quasars.  Likewise, there is excellent observational evidence for GRB
fireballs with conical geometry.  Henceforth, following standard
usage, we will interchangeably use the term ``jet'' for conical blast
waves.

As a result of relativistic beaming (``abberation''), an observer can
see only a limited portion of the blast wave with angular size $\sim
\Gamma^{-1}$, where $\Gamma$ is the bulk Lorentz factor of the blast
wave.  This relativistic beaming implies that there is no observable
distinction between a spherically expanding blast wave and a conical
blast wave (whose opening angle we denote by $\theta_j$) until the
blast wave has slowed down to $\Gamma<\theta_j^{-1}$.  However,
gamma-ray emission is expected to occur when $\Gamma$ is large,
$\Gamma \simgt 100$, and thus unless the opening angles are very
small, $\theta_j<0.01$, a conical GRB will not light up the full
celestial sphere, but only the so-called beaming fraction $f_b
=(1-\cos \theta_j)$; we note that for $\theta_j\simlt 1$, $f_b\cong
\theta_j^2/2$.

There are two consequences of relativistic beaming: (1) the true GRB
rate is $f_b^{-1}$ times larger than the observed GRB rate and (2) the
true gamma-ray energy released $E_\gamma$ is smaller than $E_{\rm
  iso}(\gamma)$ by the same factor, i.e., $E_\gamma=f_b\times E_{\rm
  iso}(\gamma)$.

In contrast to the situation during the gamma-ray burst phase, during
the afterglow phase $\Gamma$ ranges from $\sim 10$ hours after a burst
to trans-relativistic ($\Gamma\sim 1$) and even non-relativistic
values days to months after the burst.  Thus at some point during the
lifetime of the afterglow $\Gamma$ falls below $\theta_j^{-1}$ which
has a clear observational signature.  Thus multi-wavelength (X-ray,
optical and radio) afterglow observations offer us an elegant way to
measure $\theta_j$.  Here, we report an extensive analysis of
$\theta_j$ values for all well-studied GRBs.  From these values of
$\theta_j$ we are able to infer the true $\gamma$-ray energy release
of the central engines of GRBs.

The organization of the paper is as follows. First we summarize the
physics of conical afterglows followed by extraction of $\theta_j$
values from afterglow observations. We then find the surprising result
that $E_\gamma$ is tightly
clustered around $5\times 10^{50}$ erg. We end with a discussion of
the ramifications of this result.

\medskip
\noindent{\bf Conical Afterglows} 
\medskip

The temporal evolution of the afterglow emission is directly related
to the dynamics of the blast wave which in turn is influenced both by
the circumburst medium and the geometry of the explosion.  The
afterglow emission from a conical blast
wave\cite{rho97b,wkf98,pmr98,rho99,sph99} differs from that of a
spherical blast wave in two distinct ways.  First, the observer will
start noticing a deficit of emitting material when $\Gamma <
\theta_j^{-1}$.  The magnitude of this deficit, relative to that of a
spherical fireball, is proportional to the ratio of the area of the
emitting surface for a conical blast wave ($\propto \theta_j^2$) to
that of a spherically emitting blast wave ($\propto \Gamma^{-2}$).
This deficit results in the afterglow emission declining more rapidly,
relative to a spherical case, or a ``break'' in the power law decay,
$\Delta\alpha=3/4$.

The second effect that also becomes important when $\Gamma\simlt
\theta_j^{-1}$ is the spreading of the jet in the lateral
dimension\cite{rho99,sph99}. The ejecta now encounter more surrounding
matter and decelerate faster than in the spherical case.  This results
in an overall steepening ($\Delta \alpha=1.2$--1.5) of the afterglow
emission.  In the case of a laterally spreading jet, the lightcurve
evolves with $\alpha\cong p$. We note $p\sim$ 2.2--2.4 appears to fit
all well-studied afterglows (e.g., ref.~\pcite{sph99}, \pcite{fw01}).
This value is also favored by shock acceleration models\cite{gak99}.

The first claim of a jet was made for the radio afterglow of
GRB~970508, which showed deviations from the predictions of a simple
spherical adiabatic model\cite{wkf98}. However, it was the spectacular
isotropic energy release\cite{kdo+99} of GRB 990123 -- approaching the
rest mass of a neutron star -- which emphasized the possible
importance of jets in GRBs.  A case for a jet in the afterglow of this
burst was made on the basis of a sharp break ($\Delta\alpha\geq
0.7$)\cite{ftw00} in the optical afterglow and upper limits in the
radio\cite{kfs+99}.  The clearest evidence for a jet is a sharp break
over a broad range of frequencies and such a signature was seen in the
lightcurves of GRB 990510 at optical\cite{sgk+99b,hbf+99} and
radio\cite{hbf+99} wavelengths and was found to be consistent with the
X-ray\cite{psa+00} light curve.  Furthermore, the detection of
polarization\cite{clg+99,wvg+99} from this event gave further credence
to the jet hypothesis: the non-spherical geometry leads to polarized
signal, from which the geometry of the jet can be
inferred\cite{gl99,sar99}.

More recently, the identification of jets has shifted from single
frequency measurements to global model fitting of joint optical, radio
and X-ray datasets (e.g., ref.~\pcite{bsf+00}, \pcite{pk00b}). This
approach has the advantage that by simultaneously fitting all the
data, the final outcome is less sensitive to deviations in small
subsets of the data.  In addition, since the character of the
achromatic break is different above and below the peak of the
synchrotron spectrum\cite{sph99}, broad-band measurements give more
robust determinations of the jet parameters. This approach was crucial
in distinguishing the jet break for GRB\,000301C\cite{bsf+00} whose
decaying lightcurves exhibited unusual variability\cite{mbb+00}, now
attributed to microlensing\cite{gls00}.

\medskip
\noindent{\bf Determination of the Jet Opening Angles}\label{sec:geometry}
\medskip

We use the formulation of Sari, Piran \&\ Halpern\cite{sph99} to
convert the measured jet break times $t_j$ to opening angles of the
conical blast wave:
\begin{equation}
        \theta_j = 0.057 
              \Bigl({t_j \over 1\, {\rm day}}\Bigr)^{3/8} 
              \Bigl({1+z \over 2}\Bigr)^{-3/8}  
              \Bigl({E_{\rm iso}(\gamma) \over 10^{53}\,{\rm erg}}\Bigr)^{-1/8} 
              \Bigl({\eta_\gamma \over 0.2}\Bigr)^{1/8} 
              \Bigl({n \over 0.1\,{\rm cm^{-3}} }\Bigr)^{1/8},
\label{eq:theta-j}
\end{equation}
where $\eta_\gamma$ is the efficiency of the fireball in converting
the energy in the ejecta into $\gamma$ rays, and $n$ is the mean
circumburst density.  In Table~\ref{tab:Table-jet} we present a
complete sample of all GRBs with known redshifts as of December 2000.
The determinations of $t_j$ are of varying quality.  The best events
are those for which it is possible to globally model the broad-band
data within the physical framework of the relativistic jet model
(e.g., GRB 000301C, GRB 990510). For some bursts $t_j$ is inferred
from only one band (e.g., GRB 990705) and in some cases with
additional constraints from radio observations (e.g., GRB 990123).
Finally, there are some events with only upper (e.g., GRB 991208) or
lower limits on $t_j$ (e.g., GRB 971214), for which only upper or
lower limits of $\theta_j$ can be placed, respectively.

\begin{figure*}
  \centerline{\hbox{\psfig{figure=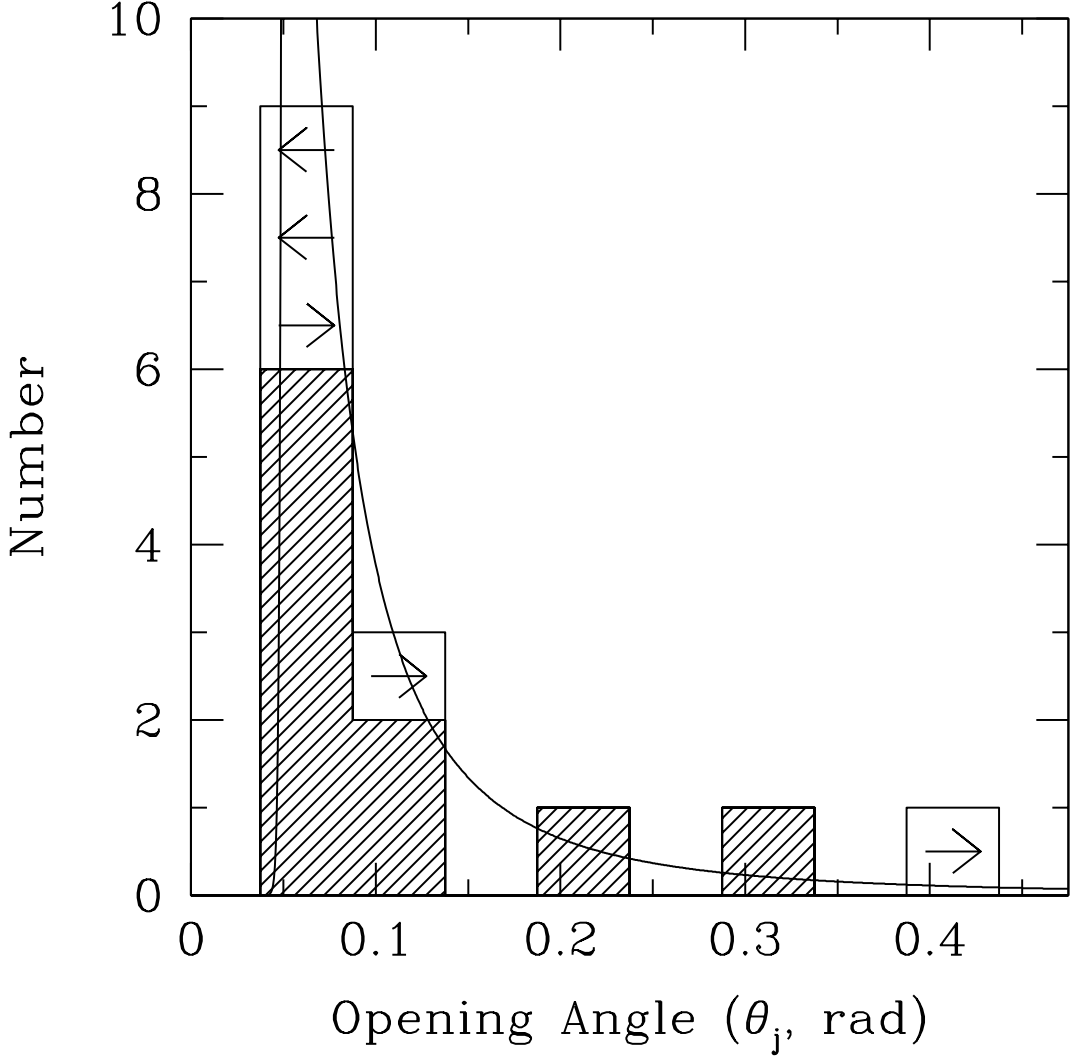,width=15cm}}}
\caption[]{\protect 
  The observed distribution of jet opening angles along with a model
  fit (line).  We assume that the observed differential distribution
  of beaming factors can be represented by two power laws: $p_{\rm
    obs}(f_b)=(f_b/f_0)^{\alpha+1}$ for $f_b<f_0$ and $p_{\rm
    obs}(f_b)=(f_b/f_0)^{\beta+1}$ for $f_b>f_0$.  Since for every
  observed burst there are $f_b^{-1}$ that are not observed, the true
  distribution is $p_{\rm true}(f_b) = f_b^{-1} p_{\rm obs}(f_b)$.
  Fitting to the data, we find the following: $\alpha$ is poorly
  constrained; $\beta=-2.77^{+0.24}_{-0.30}$; $\log f_0
  =-2.91^{+0.07}_{-0.06}$.  Thus, the true differential probability
  distribution (under the small angle approximation, $f_b\propto
  \theta_j^2$) is given by $p_{\rm true}(\theta_j)\propto
  \theta_j^{-4.54}$ with the observed distribution being $p_{\rm obs}
  \propto \theta_j^{-2.54}$.  The distribution $p_{\rm true}(f_b)$
  allows us to estimate the true correction factor, $\langle
  f_b^{-1}\rangle$ that has to be applied to the observed GRB rate in
  order to obtain the true GRB rate. We find $\langle f_b^{-1}\rangle
  = f_0^{-1}[(\beta-1)/\beta] \sim 520\pm 85$.}
\label{fig:opening}
\end{figure*}

We obtain a range in $\theta_j$ corresponding to the wide range in
$t_j$ values in Table~\ref{tab:Table-jet} (from $\simlt 1$ d to 30 d).
The derived jet angles vary from 3$^\circ$ to more than 25$^\circ$
with a strong concentration near 4$^\circ$ (Figure~\ref{fig:opening}).
It is reasonable to ask whether the observed distribution in Figure
\ref{fig:opening} suffers from selection effects.  To begin we note
that out of the 21 known optical afterglows, the light curves of only
two GRBs -- GRB 980326 (ref.~\pcite{ggv+98c}) and GRB 980519
(ref.~\pcite{jhb+00}) -- show rapid decline implying $t_j\simlt 1$ d.
Likewise, out of a sample of 10 bright X-ray afterglows observed with
the BeppoSAX satellite there is no evidence for a significant break
within 8 to 48 hours after a burst\cite{spg+00}, suggesting that $t_j
\simgt 1$\,d for these events. If we increase the sample to include
the 28 GRBs detected by BeppoSAX for which follow-up searches
(typically 8--12 hr after the burst) were made for an X-ray afterglow
we find only one unambiguous case where no afterglow was detected
(GRB\,990217; ref.~\pcite{cos00}). There are a further six cases where
a hitherto uncataloged X-ray source was detected in the GRB error
circle. In every case the X-ray source is a plausible afterglow but
lacking multi-wavelength confirmation, the afterglow identification
remains uncertain, e.g., GRB 970111, ref.~\pcite{fag+98}. From these
statistics we conclude that steep decays, $t_j\simlt 1$ d, and
therefore very narrow opening angles, $\theta_j<3^\circ$, are required
for less than ten percent of the BeppoSAX GRB sample.

There is another method to infer the existence of a population of GRBs
with extremely narrow opening angles. The beaming fraction during the
afterglow phase is $\max(\theta_j^2/2, ~\Gamma^{-2}/2)$. Thus, while
narrow-angle GRBs will be rare, their X-ray, optical, and radio
afterglows which are emitted at increasingly smaller $\Gamma$ are
accordingly less rare\cite{rho97b}.  However the current
limits\cite{pl98,gri99} of these ``burst-less'' afterglows do not
place further significant constraints on $\theta_j$.

GRBs with large opening angles do not suffer from severe beaming but
it is not easy to measure $t_j$ for such bursts.  For large $t_j$ the
afterglow emission is weak and (at optical wavelengths) the host
galaxy starts dominating\cite{hum+00,die01}. Thus optical observations
and X-ray are unlikely to yield $t_j$. Fortunately, radio observations
can and do play a crucial role, due to the long lifetime of the
afterglow in this regime.  This was the case for four out of five
wide-angle jets identified in Table~\ref{tab:Table-jet}. One notable
example is GRB\,970508 where a jet model\cite{fwk00} of the radio data
was found to be consistent with an analysis of the optical light
curves\cite{rho99}.


\medskip
\noindent{\bf The Luminosities and Energies of GRB Central Engines}
\medskip

\begin{figure*}
  \centerline{\hbox{\psfig{figure=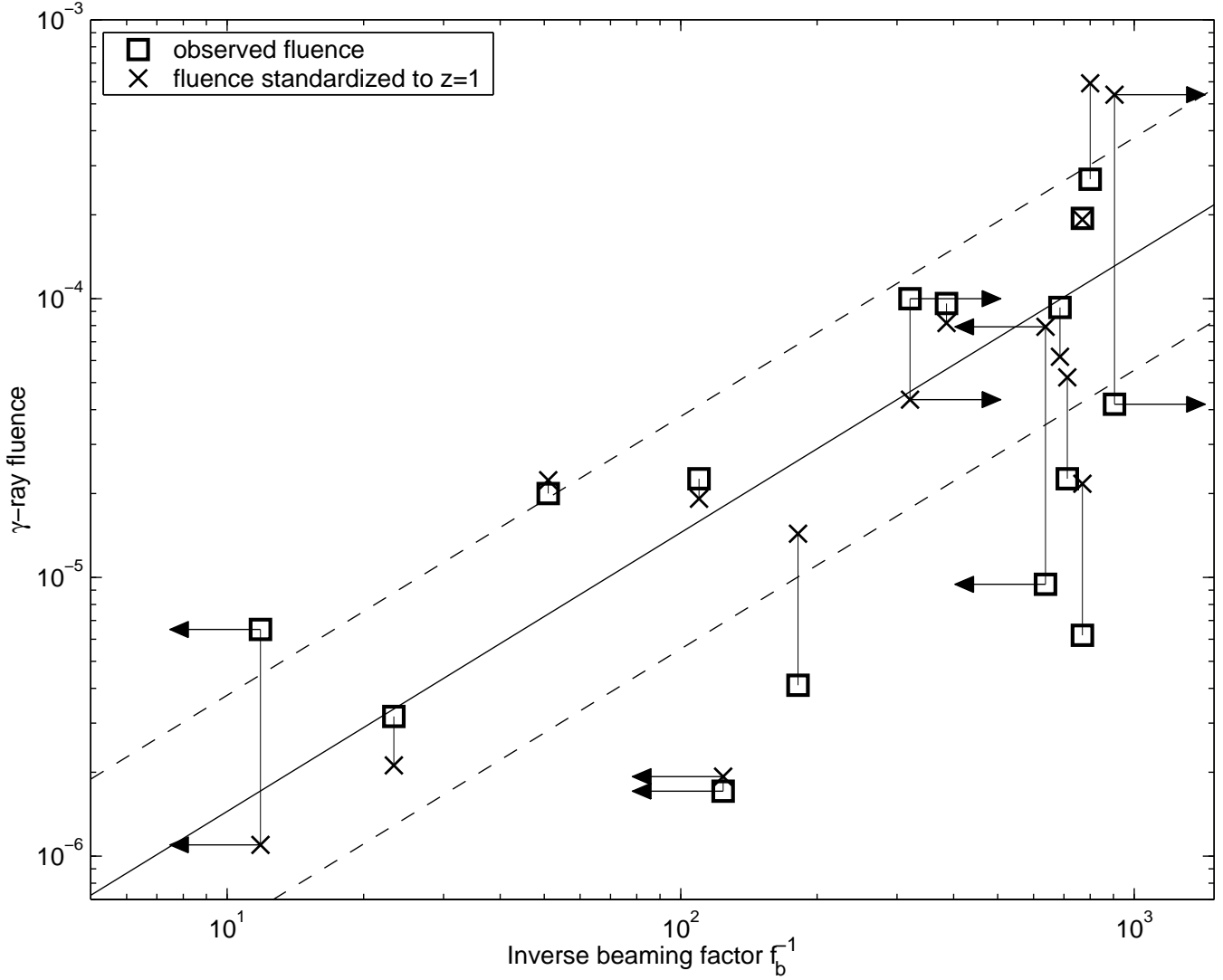,width=18cm}}}
\caption[]{\protect The gamma-ray fluence F$_\gamma$ (in units of erg
  cm$^{-2}$) plotted as a function of the inverse beaming fraction
  $f_b^{-1}$ (where $f_b= 1-\cos \theta_j \cong \theta_j^2/2$, and
  $\theta_j$ is the opening angle of the jet). A correlation is
  apparent in the sense that GRBs that have narrower jet opening
  angles are brighter (high fluence) than those that do not. A linear
  fit to these data (open squares) gives a relatively large rms
  scatter of a factor of 3.3. The correlation is improved when the
  fluences are all scaled to the same unity redshift (crosses),
  thereby removing the distance dependence. The rms scatter (dashed
  lines) of these points around a linear fit (solid line) is reduced
  to only a factor of 2.3.  This factor is marked by dashed lines
  around the linear fit. The wide variation in observed fluence, more
  than two orders of magnitude, appears to be mainly due to different
  beaming angles.}
\label{fig:theta-fluence}
\end{figure*}

\begin{figure*}
  \centerline{\hbox{\psfig{figure=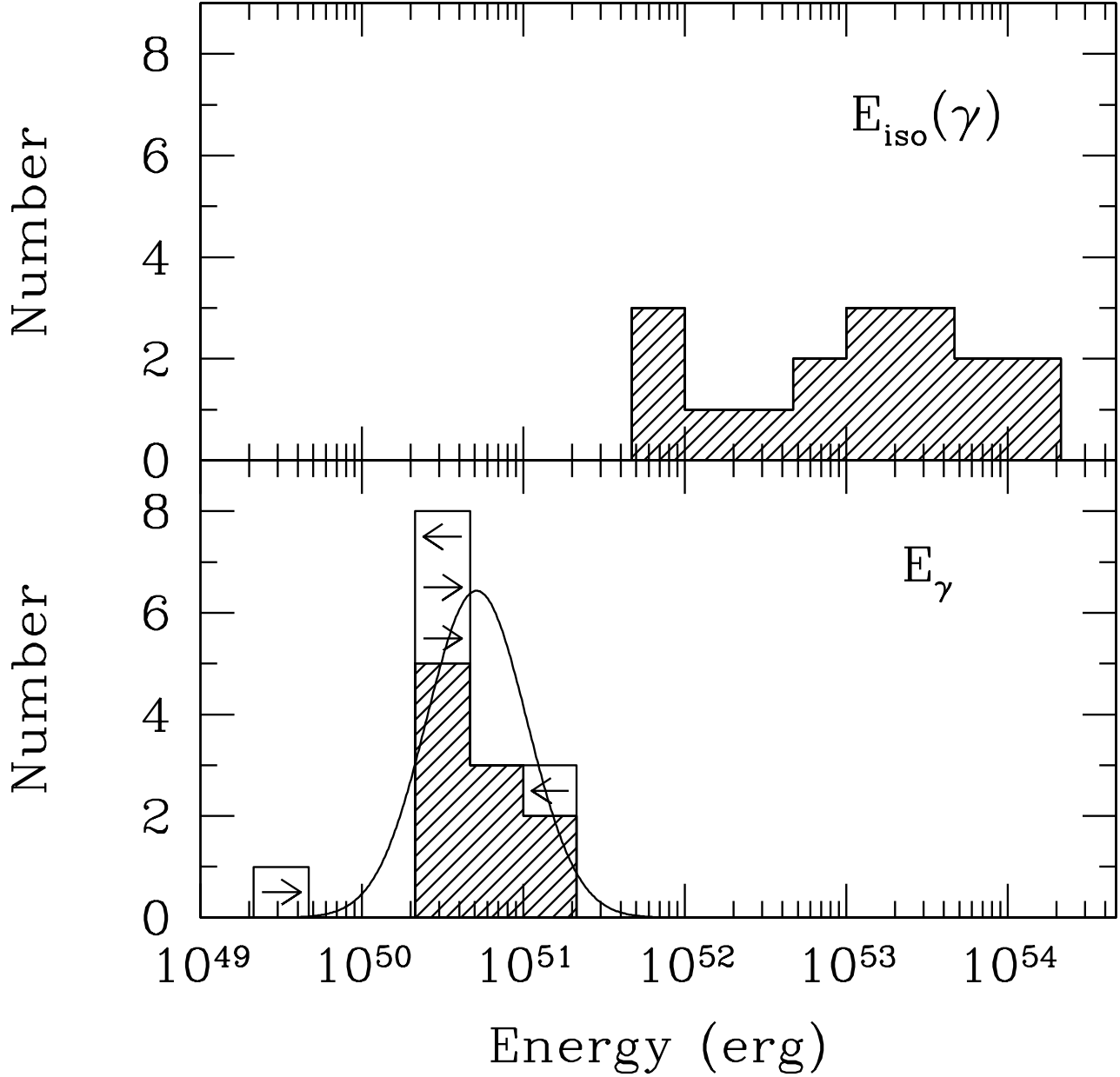,angle=0,width=18cm}}}
\caption[]{\protect 
  The distribution of the apparent isotropic $\gamma$-ray burst energy
  of GRBs with known redshifts (top) versus the geometry-corrected
  energy for those GRBs whose afterglows exhibit the signature of a
  non-isotropic outflow (bottom).  The mean isotropic equivalent
  energy $\langle E_{iso}(\gamma)\rangle$ for 17 GRBs is $110 \times
  10^{51}$ erg with a $1$-$\sigma$ spreading of a multiplicative
  factor of 6.2. In estimating the mean geometry-corrected energy
  ${\langle E_\gamma\rangle}$ we applied the Bayesian inference
  formalism\cite{rei01} and modified to handle datasets containing
  upper and lower limits\cite{rlf+01}. Arrows are plotted for five
  GRBs to indicate upper or lower limits to the geometry-corrected
  energy. The value of $\langle \log{E_\gamma}\rangle$ is
  50.71$\pm$0.10 (1$\sigma$) or equivalently, the mean
  geometry-corrected energy $\langle E_\gamma\rangle$ for 15 GRBs is
  $0.5\times 10^{51}$ erg. The standard deviation in $\log{E_\gamma}$
  is 0.31$^{+0.09}_{-0.06}$, or a $1$-$\sigma$ spread corresponding to
  a multiplicative factor of $2.0$.}
\label{fig:histo}
\end{figure*}

In Figure~\ref{fig:theta-fluence} we plot the measured fluence versus
the inferred inverse beaming factor. There appears to be a correlation
in the sense that the bursts with the largest fluence have the
narrowest opening angles.  This trend was noted earlier\cite{sph99}
albeit based on a few afterglows.  The correlation is improved when
the fluences are all scaled to the same redshift (unity), which
effectively renders it to a correlation between $E_{\rm iso}(\gamma)$
and $f_b^{-1}$.  The physical meaning of this trend is better
appreciated from Figure~\ref{fig:histo} where we find that $E_\gamma$,
the true energy released in gamma rays, is clustered around $5\times
10^{50}$ erg, with a 1$\sigma$ multiplicative factor of only two.

Figures~\ref{fig:theta-fluence} and \ref{fig:histo} suggest the
following simple scenario: the central engines of GRBs produce
approximately a similar amount of energy, and a significant part,
about $10^{51}$ erg, escapes as gamma-rays (Figure~\ref{fig:histo}).
However, for reasons not presently understood, there exists a wide
range of jet opening angles.  If so, GRBs with the narrowest opening
angles would be brighter and consequently produce the correlation seen
in Figure~\ref{fig:theta-fluence}.

The narrowness of the $E_\gamma$ distribution is surprising and has
several immediate implications.  While it is not unreasonable to
expect that the central engines produce a similar amount of energy,
$E_0$, in each explosion, there is little reason to expect that they
will produce similar gamma-ray outputs. Since the true total energy
$E_0 \equiv E_\gamma/\eta_\gamma \propto n^{1/4} \eta_\gamma^{-3/4}$
(this follows from Equation \ref{eq:theta-j}), the narrowness in the
distribution of $E_\gamma$ places restrictions on the dispersion of
$n$ and $\eta_\gamma$.

If $\eta_\gamma$ is high (close to unity) then a small dispersion in
$\eta_\gamma$ is naturally assured.  Indeed, a number of recent
papers\cite{bel00,gsw01,ks01} have argued that internal shocks under
certain conditions are very efficient at producing gamma rays
($\eta_\gamma\simgt 0.2$).  Furthermore, Guetta et al.\cite{gsw01}
argue that the very conditions that are needed to make internal shocks
efficient (a large dispersion in the distribution of the ejecta's
Lorentz factors) also produce the characteristic clustering of
spectral break energies of GRBs in the range 0.1--1 MeV.

Given a distribution in $E_\gamma$ with a full width of a factor of
four (see Figure \ref{fig:histo}), the dispersion in $n$ ($\propto
E_\gamma^4$) has to be less than two orders of magnitude.  At first
this may seem to be a weak constraint on the possible progenitors.
However, the progenitors discussed to date lie either in intergalactic
space or the halos of galaxies (ns-ns coalescence, $n\sim 10^{-6}$
cm$^{-3}$ and $n\sim 10^{-4}$ cm$^{-3}$) or in a typical disk
interstellar medium (ISM) or dense ISM (collapsar; $n\sim 1$ cm$^{-3}$
and $n\sim 10^2$ cm$^{-3}$, respectively).  Therefore our results
limit the diversity of GRB environments, and specifically requires
that the long-duration class of GRB events happen in only one of these
environments.  Furthermore, we note that winds of massive stars would
produce a density of a few atoms cm$^{-3}$ for $t_j\simgt 1$ d. In
scenarios where there are two types of GRBs\cite{cl00}, the ones that
do not go off in stellar-wind-stratified media must reside within the
disk of their host galaxy rather than in galaxy halos or the
intergalactic medium.  Indeed, the distribution of GRBs within their
host galaxies is consistent with a disk population\cite{bkd00}.
Likewise, broad-band modeling of GRB
afterglows\cite{bsf+00,pk00b,fwk00} give estimates of gas densities
consistent with disks, justifying our normalization of $n$ in Equation
\ref{eq:theta-j}. We conclude that the progenitors of long duration
GRBs likely come from one type of progenitor.

Finally, the narrowness of the $E_\gamma$ distribution requires that
the brightness of the $\gamma$-ray beam be roughly uniform from the
center to the edge. This is contrary to models\cite{kp00a} in which
large intensity variations within the conical blast wave are invoked
in order to explain the wide dispersion of peak luminosities. We find
that most of the dispersion in the luminosity is due to the diversity
in opening angles.

The mean value of $E_{\gamma}$ is $5\times 10^{50}$ erg
(Figure~\ref{fig:histo}). If we accept the conclusions of Guetta et
al.\cite{gsw01} (see above), then $\eta_\gamma\sim 0.2$ and we then
derive $E_0\sim 3\times 10^{51}$ erg.  Of course, $E_0$ is sensitive,
in addition to the adopted value of $\eta_\gamma$, to the overall
scaling, i.e., the numerical coefficient of Equation \ref{eq:theta-j}.
For example, the estimate of Rhoads\cite{rho99}, based on a different
assumption for the sideways expansion speed, has a coefficient smaller
by a factor of six than our Equation \ref{eq:theta-j}.

Fortunately, GRB 970508 allows us to directly determine the energy
scale. The radio afterglow of this GRB lasted long enough (400 d) that
the blast wave was non-relativistic, thereby allowing determination of
the total energy\cite{fwk00} independent of relativistic beaming.
Table \ref{tab:Table-jet} shows that this burst has one of the lowest
energies, although it is only $1\sigma$ away from the mean (if the
energy distribution is assumed to be log normal). The agreement
between these two entirely different approaches is remarkably good and
gives some support to our choice of the numerical coefficient and
normalization of Equation~\ref{eq:theta-j}.

Freedman \&\ Waxman\cite{fw01} and Kumar\cite{kumar00} have suggested
an elegant way to estimate the energy in the afterglow phase based on
X-ray observations.  This method yields the $\epsilon_e \varepsilon_a$
where $\varepsilon_a$ is the energy of the blast wave per steradian
and $\epsilon_e$ is the fraction of energy in the shocked electrons.
This estimate is independent of the ambient density.  If $\epsilon_e$
is high and relatively constant (analogous to the situation with
$\eta_\gamma$) then $E_0$ can be estimated provided $f_b$ is known.
It is of interest to note that the ratio $\varepsilon_a/E_{\rm
  iso}(\gamma)$ is nearly constant\cite{fw01}, suggesting that likely
both $\epsilon_e$ and $\eta_\gamma$ are narrowly distributed.

Applying our determinations of $f_b$ to the sample of ref.
\pcite{fw01} (a total of six common GRBs) we obtain $E_a=(2.7\pm
1.4)\times 10^{50}$ erg.  Within the limitations of the small sample,
the distribution appears to be clustered and the results are in
agreement with our findings (Figure~\ref{fig:histo}).  The principal
advantage of our method is that the events are always identified in
the $\gamma$-ray band, whereas X-ray observations are available for
only a minority of cases. Furthermore, the X-ray afterglow technique
ignores the effects of inverse Compton scattering effects\cite{se00},
and is therefore sensitive to the poorly known strengths of magnetic
fields in strong shocks.

\medskip
\noindent{\bf GRBs and SNe}
\medskip

Above we find that the mean total energy of GRBs is $E_0\sim 3\times
10^{51}$ erg. This energy is only slightly larger than the typical
$10^{51}$ erg of electromagnetic and kinetic energy yield of ordinary
supernovae (Ia, Ibc, II). This reduced energy budget raises the
possibility that GRBs are the result of the formation of neutron
stars\cite{wyh+00}, albeit with special properties\cite{uso92}, and
does not necessarily require black holes.  The mystery about GRBs is
no longer in understanding their supposedly extraordinary energy
budget but in explaining why the ejecta of GRBs have such a high
Lorentz factor.


We note however that there are at least two possible exceptions to the
tight clustering of jet energy. {\it (1)} If SN 1998bw is associated
with a GRB\cite{gvv+98,kfw+98} then $E_{\rm iso}(\gamma) \sim 7\times
10^{47}$ erg. However, Kulkarni et al.\cite{kfw+98} have argued that
the extraordinarily bright radio emission from this SN requires
$\simgt\,10^{50}$ erg of the explosion energy to be in the form of
mildly relativistic ejecta ($\Gamma \sim$ few). {\it (2)} Bloom et
al.\cite{bkd+99} identify the late time red bump in the rapidly
decaying event GRB 980326 ($t_j<0.55$ d; ref. \pcite{ggv+98c}) with an
underlying SN. If so, the inferred redshift $z\sim 1$ and $E_\gamma <
7\times 10^{49}$ erg.  Unfortunately, the radio observations are not
sensitive enough to place meaningful constraints on the amount of
energy in mildly relativistic ejecta.  In both cases, the true energy
release could be closer to $E_0$ but this energy could be primarily in
mildly relativistic ejecta.  Careful observations (especially X-ray
and radio) of SNe may uncover significant numbers of such ``failed''
GRBs.

\medskip
\noindent{\bf Beaming Fraction and the GRB Rate}
\medskip

Since conical fireballs are visible to only a fraction, $f_b$, of
observers, the true GRB rate, $R_t=\langle f_b^{-1}\rangle R_{\rm
  obs}$, where $R_{\rm obs}$ is the observed GRB rate and $\langle
f_b^{-1}\rangle$ is the harmonic mean of the beaming fractions. We
find $\langle f_b^{-1}\rangle\sim 500$ (see caption to
Figure~\ref{fig:opening}). The formal uncertainty in this estimate is
only 16\% but systematic uncertainties related our choice of the
numerical coefficient and normalization of Equation~\ref{eq:theta-j}.
make this estimate accurate to a factor of two.

Estimates\cite{wbbn98,sch01,klk+00} of the local {\it observed} rate
of GRBs give values of $R_{\rm obs}(z=0)$ ranging from 0.2 to 0.7
Gpc$^{-3}$ yr$^{-1}$. The rate is uncertain because it is not known
how the GRB rate evolves with redshift. We adopt a value $R_{\rm
  obs}(z=0)=0.5$ Gpc$^{-3}$ yr$^{-1}$ as in Ref. \pcite{sch01}. The
true rate is $R_t(z=0) \sim 250 $ Gpc$^{-3}$ yr$^{-1}$, which should
be compared with the estimated rate\cite{phi91} of neutron star
coalescence, $R_{\rm c}(z=0)\sim 80$ Gpc$^{-3}$ yr$^{-1}$ and the
estimated rate\cite{phi91} of type Ibc SN, $R_{\rm Ibc}\sim 6\times
10^4$ Gpc$^{-3}$ yr$^{-1}$.  Clearly, the collapsar scenario is
capable of easily supplying a sufficient number of progenitors
(including failed GRBs). Within the uncertainties of the estimates,
the coalescence scenario is also (barely) capable of providing
sufficient progenitors.

\medskip
\noindent{\bf Assumptions, Uncertainties and Caveats}
\medskip

Our derivation of the jet opening angle is based on
Equation~\ref{eq:theta-j} which makes two implicit assumptions.
First, we assume that GRBs explode in a constant density medium and
that any sharp break in the afterglow ($\Delta\alpha \simgt 0.75$) is
attributed to a combination of the observer viewing beyond the edge of
the conical jet and sideways expansion.  Second, we assume that the
conical blast wave maintains a fixed opening angle right from the GRB
phase until $\Gamma$ approaches $\theta_j^{-1}$.  The latter
assumption imposes strict conditions on the working of the central
engine. Specifically, the ejecta have to be approximately uniform
across the entire opening angle in the gamma-ray phase and the bulk of
the explosive energy in the afterglow phase must have a single bulk
Lorentz factor.

The origin of the observed break is currently a matter of considerable
theoretical debate\cite{pm99,kp00b,wl00}.  The uncertainty is driven
by the as yet unclear hydrodynamics of sideways expansion.  Some
authors\cite{pm99,pk00} argue that transition is very smooth, and is
completed in one decade of the break time for a constant density
circumburst medium but takes two decades for the $n(r)\propto r^{-2}$
circumburst medium.  Others\cite{sph99} argue that several
uncertainties in these calculations make this conclusion premature.

Furthermore, the analysis of the afterglow lightcurves does not yield
$\phi$, the angle between the line of sight to the observer and the
principal axis of the jet.  In general one may expect some dependence
of $t_j$ on $\phi$ and thus $E_\gamma$ distribution should broaden
even if $E_0$, $\eta_\gamma$ and $n$ were constants.

However, the narrowness of the $E_\gamma$ distribution shown in
Figure~\ref{fig:histo} provides empirical support for our assumptions.
Furthermore, as noted earlier, where high quality observations are
available (e.g., GRB 990510 and GRB 000301C) the breaks are found to
be quite sharp and the inferred ambient density is $\sim 1$ cm$^{-3}$.

Finally, several other mechanisms have been proposed to produce steep
declines in the afterglow light curves: {\it (i)} a sudden drop in the
external density\cite{kp00b}, {\it (ii)} transition from relativistic
to non-relativistic regime\cite{wdl00} due to expansion in a dense
circumburst medium, and {\it (iii)} a break in the power-law
distribution of radiating electrons\cite{lc00}.  These models have not
been systematically compared against well studied afterglows and thus
remain at a level of suggestions. Model {\it (ii)} can be rejected
because the expected centimeter wave attenuation due to free-free
absorption is not seen.  Model {\it (iii)} is unable\cite{lc00} to
explain the broad-band achromatic breaks and will therefore fail to
account for the early time low frequency emission.  We conclude that
at this stage the simple jet model which we have adopted provides a
consistent and adequate description of the observations.

Our understanding of gamma-ray bursts has increased dramatically over
the past four years.  For nearly three decades these objects were
considered so enigmatic that expectations of their distance ranged
from local to cosmological scales. In the BATSE era, prior to the
discovery of the afterglow phenomenon, the standard assumption was
that GRBs possessed fixed peak luminosities.  As more and more
redshifts were obtained, the isotropic equivalent energy record
increased, eventually reaching the rest mass energy of neutron stars.
The standard candle hypothesis was consequently abandoned. It is
remarkable that with a more detailed understanding of the afterglow we
are able to infer the energy release in these bursts and find that
GRBs are ``standard candles'' in some sense.

We have deduced the distribution of the opening angles of GRB jets and
empirically uncovered a key clue, namely the total energy release and
its approximate constancy, but we are still left with three
significant mysteries.  First, we do not know what physical mechanism
results in such a wide variation in the opening angles of the jets.
Second, the similarity of the energy release in GRBs and ordinary
supernovae is puzzling.  This coincidence is all the more remarkable
considering the diversity of the progenitors and likely differing
collapse mechanisms in these various classes of explosions.  Third, we
do not understand why in GRBs, the explosion energy couples only to
$10^{-5}\,M_\odot$ of the exploding star and thereby produce ejecta
with high Lorentz factor.  Fortunately, new missions (HETE-2 and
SWIFT) with their vastly increased GRB localization rates will provide
empirical data which may help solve these mysteries.

\clearpage

\begin{acknowledge}
  Our research is supported by NASA and NSF. JSB thanks the Fannie \&\ 
  John Hertz Foundation for their generous support, AD holds a
  Millikan Postdoctoral Fellowship in Experimental Physics, TJG holds
  a Fairchild Foundation Postdoctoral Fellowship in Observational
  Astronomy, DER holds a Hubble Fellowship and RS holds a Fairchild
  Foundation Senior Fellowship in Theoretical Astrophysics.  The
  National Radio Astronomy Observatory is a facility of the National
  Science Foundation operated under cooperative agreement by
  Associated Universities, Inc. We thank the staff at the Keck and
  Palomar Observatoroes for their expert help during many observing
  runs. The W. M. Keck Observatory is operated by the California
  Association for Research in Astronomy, a scientific partnership
  among the California Institute of Technology, the University of
  California and the National Aeronautics and Space Administration.
  It was made possible by the generous financial support of the W. M.
  Keck Foundation.
\end{acknowledge}

\clearpage

\clearpage

\newpage 
\begin{table}
\begin{tabular}{|l|r|l|r|r|r|r|r|r|r|}
\hline
GRB    &$F_\gamma$ & $z$ & $d_L$  &$E_{\rm iso}(\gamma)$ & $t_j$    & $\theta_j$ &
$E_\gamma$ & Refs. & Note\\  \hline \hline
970228 & 11.0  &  0.695  & 1.4  &  22.4     &          &       &      &  & N\\ \hline
970508 & 3.17  &  0.835  & 1.8  &  5.46     &  25      & 0.293 & 0.234& \pcite{fwk00} & R\\   \hline
970828 & 96.0  &  0.958  & 2.1  &  220      &  2.2     & 0.072 & 0.575 & \pcite{dfk+00} & X\\\hline
971214 & 9.44  &  3.418  & 9.9  &  211      & $>$ 2.5 & $>$ 0.056 & $>$ 0.333 & \pcite{kdr+98} & O\\ \hline
980613 & 1.71  &  1.096  & 2.5  &  5.67     & $>$ 3.1  & $>$ 0.127 & $>$ 0.045 & \pcite{hf98}  & O\\ \hline
980703 & 22.6  &  0.966  & 2.1  &  60.1     & 7.5      & 0.135 & 0.544 & \pcite{ber01} & B\\   \hline
990123 & 268   &  1.600  & 3.9  &  1440     & 2.04     & 0.050 & 1.80 & \pcite{kdo+99} & O\\ \hline
990506 & 194   &  1.30   & 3.0  &  854      &          &       &      &  & N\\ \hline
990510 & 22.6  &  1.619  & 4.0  &  176      & 1.20     & 0.053 & 0.248& \pcite{hbf+99} & B\\ \hline
990705 & 93    &  0.84   & 1.8  &  270      & $\sim$1  & 0.054 & 0.389& \pcite{mpp+00} & O\\ \hline
990712 & 6.5   &  0.433  & 0.8  &  5.27     & $>$ 47.7 & $>$ 0.411 &$>$ 0.445& \pcite{fvhp00} & O\\ \hline
991208 & 100   &  0.706  & 1.4  &  147      & $<$ 2.1  & $<$ 0.079 &$<$ 0.455  & \pcite{jhp+99} & D\\  \hline
991216 & 194   &  1.02   & 2.3  &  535      & 1.2      & 0.051 & 0.695 & \pcite{hum+00} & O\\ \hline
000131 & 41.8  &  4.500  & 13.7 &  1160     & $<$ 3.5  & $<$ 0.047 &$<$ 1.30 & \pcite{ahp+00}  & D\\      \hline
000301C& 4.1   &  2.034  & 5.3  &  46.4     & 5.5      & 0.105 & 0.256& \pcite{bsf+00} & B\\   \hline
000418 & 20.0  &  1.119  & 2.5  &  82.0     & 25       & 0.198 & 1.60 & \pcite{die01}  & B\\   \hline
000926 & 6.2   &  2.037  & 5.3  &  297      & 1.45     & 0.051 & 0.379& \pcite{phg+00} & O\\ \hline
\end{tabular}
\newpage
\caption{{\bf Jet Break Times and Energetics.}\label{tab:Table-jet}
  {\protect\small The gamma-ray fluences \protect ($F_\gamma$), given
    in units of \protect 10$^{-6}$ erg cm$^{-3}$, are from a diverse
    collection of instruments. The best determinations of energy
    fluence are from the \protect {\it Burst and Transient Experiment} (BATSE)
    on the \protect {\it Compton Gamma-Ray Observatory} (CGRO). Most of the
    GRBs (10), prior to the de-orbit of CGRO on 2000 May 26, are BATSE
    bursts. In these cases we used fits to BATSE data which were
    integrated over the energy range from 20 to 2000 keV.  For the
    remainder of the events we used the fluence as determined from the
    Gamma-Ray Burst Monitor on the BeppoSAX satellite (40-700 keV), or
    fluences (25-1000 keV or 25-100 keV) from the Interplanetary
    Network of satellites \protect ({\it Ulysses}, KONUS, and {\it
      NEAR})\cite{blo01b}.  The luminosity distance ($d_L$) is given
    in units of \protect $10^{28}$ cm.  It was calculated from the observed
    redshift ($z$), and adopting cosmological parameters of
    \protect H$_\circ$=65 km s$^{-1}$ Mpc$^{-1}$, $\Omega_M$=0.3, and
    \protect $\Lambda_\circ$=0.7.  Other realistic cosmologies were tried but
    they did not fundamentally change our conclusions. The isotropic
    \protect $\gamma$-ray energies ($E_{\rm iso}(\gamma)$), given in units of
    10$^{51}$ erg, have been ``k-corrected'' such that all energy
    estimates are referenced to the same 20-2000 keV co-moving
    bandpass\cite{blo01b}. Although these order-of-unity corrections
    affect individual determinations of $E_{\rm iso}(\gamma)$, they do
    not affect our results derived from the sample as a whole. The jet
    break times ($t_j$), given in days, are taken from the literature.
    The notes and the references in the table indicate how $t_j$ was
    determined. The strongest evidence for collimated outflows come
    from GRBs with achromatic breaks in their broad-band light curves
    (B). In most cases such multi-frequency datasets are not
    available, so there is a second class of events with breaks
    determined primarily from radio (R), optical (O), or X-ray (X)
    data. We include here a number of events for which no break was
    observed, yielding only lower limits of $t_j$. For some GRBs the
    steep decline of the light curve, indicating a jet geometry, is
    already fully manifest at the time of the first measurement. In
    these cases (D) we have only an upper limit on $t_j$. The final
    group of GRBs are those for which $t_j$ cannot be determined (N),
    owing to complications in the light curve such as the presence of
    a supernova signature (i.e., GRB\,970828), or the lack of
    sufficient data. The beaming-corrected gamma-ray energy
    ($E_\gamma$), given in units of 10$^{51}$ erg, was calculated by
    applying the geometric correction factor $f_b$ to \protect $E_{\rm
      iso}(\gamma)$. }}
\end{table}

\end{document}